\documentclass[10pt,conference,letter]{ctextemp_IEEEtran}
\usepackage{amsfonts}
\IEEEoverridecommandlockouts
\usepackage{graphicx}
\usepackage{epsfig,epsf}
\usepackage{amsmath,amssymb,amsfonts}
\usepackage{color,cite}

\long\def\comment#1{}

\newfont{\bbb}{msbm10 scaled 700}

\newfont{\bb}{msbm10 scaled 1100}

\begin{document}

\title{\huge Multiple Unicast Capacity of 2-Source 2-Sink  Networks}

\author{\authorblockN{Chenwei Wang, Tiangao Gou and Syed A. Jafar}
EECS Dept., University of California, Irvine, Irvine, CA 92697 \\
Email: \{chenweiw, tgou, syed\}@uci.edu}

\maketitle

\begin{abstract}

We study the sum capacity of multiple unicasts in wired and wireless
multihop networks. With $2$ source nodes and $2$ sink nodes, there
are a total of $4$ independent unicast sessions (messages), one from
each source to each sink node (this setting is also known as an $X$
network). For wired networks with arbitrary connectivity, the sum
capacity is achieved simply by routing. For wireless networks, we
explore the degrees of freedom (DoF) of  multihop $X$ networks with
a layered structure, allowing arbitrary number of hops,  and
arbitrary connectivity within each hop. For the case when there are
no more than two relay nodes in each layer, the DoF can only take
values $1,\frac{4}{3}, \frac{3}{2}$ or $2$, based on the
connectivity of the network, for almost all values of channel
coefficients. When there are arbitrary number of relays in each
layer, the DoF can also take the value $\frac{5}{3}$. Achievability
schemes incorporate linear forwarding,  interference alignment and
aligned interference neutralization principles. Information
theoretic converse arguments specialized for the connectivity of the
network are constructed based on the  intuition from linear
dimension counting arguments.
\end{abstract}

\section{Introduction} \label{sec:intro}
Capacity characterization for multiple unicasts is one of the most
important problems in network information theory. Optimal
interference management principles are essential to the multiple
unicast problem, both in the wireless setting where interference
among concurrent transmissions is an unavoidable property of the
propagation medium, as well as in the wired network setting where
inter-session network coding gives rise to interference among
multiple flows. The study of multiple unicast networks has produced
many powerful  ideas that embrace interference --  such as network
coding and interference alignment -- and  that have shown that the
capacity limits can be much higher than possible with conventional
interference avoidance approaches that do not allow the mixing of
flows,  such as routing for wired networks and TDMA/FDMA for
wireless networks. The idea of interference alignment has been
applied primarily to \emph{single hop} wireless networks, where it
has significantly advanced the understanding of signal dimensions in
the form of degrees of freedom (DoF) characterizations. Network
coding principles are most well understood in the multicast setting
where all messages are desired by all destinations. With a few
notable exceptions (including most recently, \cite{Gou_Jafar_222,
Shomorony_Avestimehr, Cai_Letaief_Fan_Feng}), the problem of
multiple unicasts over multiple hops remains wide open for both
wired and wireless networks. In this work our goal is to make
progress on this problem, by characterizing the sum capacity and
degrees of freedom of multiple unicasts over wired and wireless
layered multihop networks, respectively.



\subsection{Problem Description}
Consider
a communication network with $M$ distributed source nodes $s_1, s_2,
\cdots, s_M$, and $N$ distributed sink nodes $d_1, d_2, \cdots,
d_N$. A total of $MN$ independent unicast sessions  are possible in
this network, one for each source-destination pair. We are
interested in the multiple unicast capacity, which we define as the
maximum possible sum-rate of all $MN$ unicast sessions as they
simultaneously flow through the network. In more standard
information-theoretic terminology, we have $MN$ independent messages
$W_{mn}, 1\leq m\leq M, 1\leq n\leq N$, with messages $W_{mn}$
originating at source $s_m$ and intended for sink $d_n$, and we wish
to find the sum-rate capacity  of  these messages. Borrowing the
corresponding nomenclature from single hop wireless networks
\cite{Jafar_Shamai, Cadambe_Jafar_X}, we refer to the setting
defined above as the $M\times N$ user $X$ network.

Aside from its significance as the original setting for interference
alignment \cite{MMK, Jafar_Shamai}, an $X$ network is interesting
because the sum capacity of an $X$ network measures the total amount
of information that can flow through the network between a \emph{set
of distributed sources} and a \emph{set of distributed destinations}
without restricting the associations between source-destination
pairs. Since each source has an independent message for each
destination, all paths that go through the network can carry desired
information. However, the total amount of information between the
set of sources and the set of destinations is, in general, different
from the min-cut between the set of sources and the set of
destinations, because of the assumption of \emph{distributed}
sources and \emph{distributed} destinations, i.e., the sources
cannot share messages and destinations cannot jointly process the
received signals. For instance, the $2\times 2$ user $X$ network in
the single hop wireless setting is shown to have DoF = $4/3$ in
\cite{Jafar_Shamai}, while the  DoF
 min-cut outer bound is $2$.

As described above, the distributed nature of sources and
destinations and the presence of a desired message from each source
to each destination are the defining features of the $X$ network
setting. The network between the sources and destinations, can be
wired or wireless, single or multiple hop. In this work, we will
study two different kinds of $X$ networks.
\begin{enumerate}
\item {Wired $X$ network}: We consider this network in the general setting, i.e., we allow any number of source nodes, any number of destination nodes, any number of hops, and arbitrary network graph topologies comprised of orthogonal noise-free links. Our goal is to characterize the sum-capacity.
\item{Layered Wireless $X$ network:} Such a network is illustrated in Fig.\ref{fig:system_model}. As shown in the figure, we restrict attention to the $M=2, ~N=2$ multihop wireless setting with a layered structure, i.e., a multihop wireless $X$ network, with arbitrary number of  layers (hops), an arbitrary number of relay nodes in each layer, and arbitrary connectivity within each hop. Because this is the wireless setting, it incorporates both interference and broadcast features of wireless propagation. Our goal is to characterize the sum DoF.
\end{enumerate}
\subsection{Summary of Contribution}
The wired $X$ network sum-capacity bears a surprisingly simple
solution. \emph{The sum-capacity is equal to the min-cut separating all sources from all destinations, and is achieved simply by routing.}
There is no need for interference alignment and there is no need for
either intra-session or inter-session network coding.
Since the proof is exceedingly simple, we will describe it
here.

Suppose we allow all sources to share all messages, and we allow all
destinations to share all their received signals. Then we have
essentially a single source, single destination network. We know
that the min-cut bound is achievable for this network and a routing
solution can be found by the Ford Fulkerson algorithm. Since only
routing is needed, there is no mixing of information, i.e., there is
no need for cooperation among source nodes or among the destination
nodes. Thus, the min-cut is also achievable in the wired $X$ network
with distributed sources and destinations.

The main focus of this paper  is on the layered multihop wireless
$X$ network. Here we proceed in two steps. First, for the case that
the number of relay nodes in each layer is no more than 2, we
provide an explicit enumeration of all possible network connectivity
patterns along with their associated DoF characterizations (in the
almost surely sense). In particular, we find that the DoF can only
take values $1, \frac{4}{3}, \frac{3}{2}, 2$. Next we allow
arbitrary number of relays in each layer and show that here, in
addition to networks with DoF values $1, \frac{4}{3}, \frac{3}{2},
2$, there exist networks with DoF $= \frac{5}{3}$. Further, these
are the only multiple unicast DoF values possible for all
connectivity patterns in  a 2-source 2-sink layered multihop
wireless network (for almost all values of channel coefficients). In
establishing these results, non-trivial achievability arguments make
use of the aligned interference neutralization concept introduced
earlier in \cite{Gou_Jafar_222}. Non-trivial outer bounds are also
needed, e.g., for the DoF = $\frac{5}{3}$ case. The intuition for
the information theoretic outer bounds is obtained from linear
dimension counting arguments.

It is interesting to contrast the $2\times 2$  $X$ network with the
2 user interference network, since the only difference between the
two settings is in the message sets, i.e., both settings can be
defined for the same physical network. While the $X$ network has 4
independent messages, the interference network has only 2
independent messages. In the \emph{one-hop wireless} setting, the
$X$ network is much more interesting than an interference network
from a DoF perspective, because the $X$ network requires
interference alignment, whereas orthogonal access is DoF-optimal for
the interference network. In the \emph{multi-hop wired} setting, the
opposite is true. The interference network is interesting because it
creates opportunities for network coding (e.g., the famous butterfly
network), but the $X$ network, as explained earlier in this section,
achieves sum-capacity through simple orthogonal access (routing),
i.e., requiring neither interference alignment nor network coding.
Finally, in the \emph{layered multi-hop wireless} setting, as it
turns out, neither the interference network, nor the $X$ network
setting is trivial. The DoF of the layered multi-hop interference
network are characterized in \cite{Shomorony_Avestimehr} and are
shown to only take values $1, 3/2$ and $2$. We show here that the
layered multi-hop $X$ network setting presents an even richer
picture and gives rise to DoF values $1, 4/3, 3/2, 5/3$ and $2$. In
both cases, both achievability and converse arguments are
non-trivial. For instance, the 2 hop layered network with 2 relays
makes use of the  idea of aligned interference neutralization,
originally introduced in \cite{Gou_Jafar_222}, whether it is the
interference network or the $X$ network. In addition, the layered
multihop $X$ network gives rise to other cases where aligned
interference neutralization is needed, such as the network with
$5/3$ DoF.

\section{System Model and Definitions}

\begin{figure}[!h] \vspace{-0.15in}\centering
\includegraphics[width=3.3in]{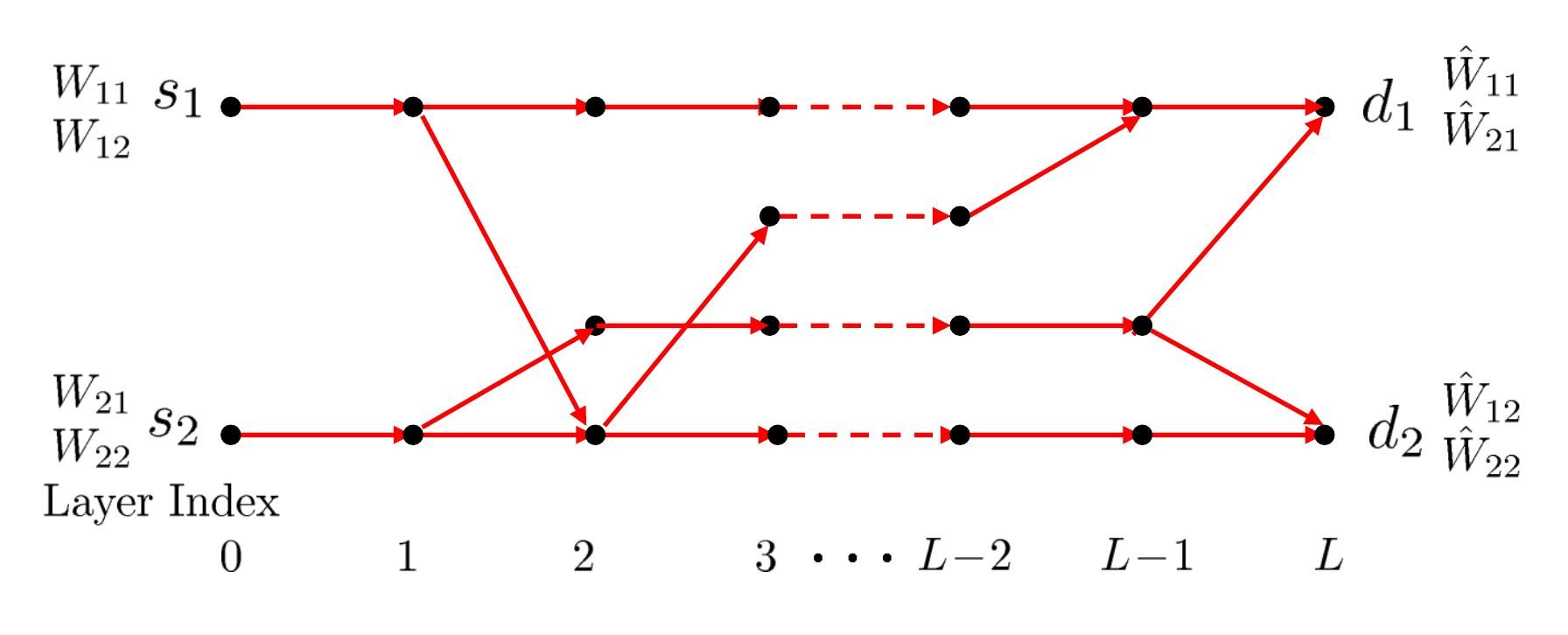}\vspace{-0.15in}
\caption{The Layered Wireless $X$ Network}\vspace{-0.1in}
\label{fig:system_model}
\end{figure}

The multihop wireless $X$ network we consider in this paper consists
of two sources $s_1, s_2$, two destinations $d_1, d_2$ and multiple
relay nodes between sources and destinations. Each node has one
antenna. There are a total of four independent messages
in this network, i.e., source $s_m$ wants to send the message
$W_{mn}$ to the destination $d_n$ where $m,n\in\{1,2\}$. We can use
a directed graph $G=(V,E)$ to characterize the network topology,
where $V$ and $E$ are the sets of nodes and edges, respectively.
Such an example is shown in Fig.\ref{fig:system_model}. The network
has a layered structure. Specifically, for a $L$-hop network,  the two
sources are at layer 0, the two destinations are at the layer $L$,
and the relay nodes at the $l^{th}~(1\leq l\leq L-1)$ layer can only
receive signals sent from the nodes at the $(l-1)^{th}$ layer, and
only transmit to nodes at the $(l+1)^{th}$ layer. In other words, in
the graph there are only edges between nodes in adjacent layers.
With the layered assumption, we consider an arbitrarily connected
network, in the sense that in any two adjacent layers, each node at
$l^{th}$ layer can be arbitrarily connected to the nodes in
${(l+1)}^{th}$ layer. We also assume that every relay node belongs to
at least one directed path from  at least one source to at least one destination, because otherwise it can
be removed without decreasing the capacity region of the network.

We denote the $i^{th}$ node in the $l^{th}$ layer as $v_i^l$. The
channel coefficient associated with the edge from the node $v_i^{l}$
to the node $v_j^{l+1}$ is denoted as $H_{v^{l+1}_jv^{l}_i}$. We
assume that the channel coefficients are independently drawn from
continuous distributions and once drawn, they remain constant during
the entire transmission. We also assume that global channel
knowledge is available at all nodes. At time index
$t\in\mathbb{Z}_+$, each node $v^{l}_i$ (except the two
destinations) transmits a complex-valued signal $X_{v^{l}_i}(t)$,
which satisfies an average power constraint
$\frac{1}{T}\sum_{t=1}^T\mathbb{E}[|X_{v^{l}_i}(t)|^2]\leq P$, for
$T$ channel uses. The signal received at the node $v^{l+1}_j$ at
time $t$ is given by
\begin{eqnarray}
\vspace{-0.125in}\mbox{$ Y_{v^{l+1}_j}(t)=\sum_{v^{l}_i\in
V^{l+1}_j} H_{v^{l+1}_j v^l_i}X_{v^{l}_i}(t)+Z_{v^{l+1}_j}(t)$}
\end{eqnarray}
where $V^{l+1}_j$ is the set of the nodes connected to $v^{l+1}_j$
at the $l^{th}$ layer, and $Z_{v^{l+1}_j}(t)$ is the i.i.d. additive
circularly symmetric complex Gaussian noise with zero-mean
unit-variance at the node $v^{l+1}_j$.

The capacity region $\mathcal {C}(\rho)$ of this network is the set
of achievable rate tuples
$R(\rho)=(R_{W_{11}}(\rho),R_{W_{12}}(\rho),R_{W_{21}}(\rho),R_{W_{22}}(\rho))$
where $\rho$ is the SNR, such that each user can simultaneously
decode its desired messages with arbitrarily small error
probability. The maximum sum rate of this channel is defined as
$R_{\textrm{sum}}(\rho)=\max_{R(\rho)\in\mathcal
{C}(\rho)}\sum_{m=1}^2\sum_{n=1}^2 R_{W_{mn}}(\rho)$. The capacity
in the high SNR regime can be characterized through DoF, i.e., DoF
$= \lim_{\rho\rightarrow \infty}R_{\textrm{sum}}(\rho)/\log\rho$.
For simplicity, we use $d_{mn}$ to denote the number of DoF
associated with the message $W_{mn}$. Note that we use the notation
$o(x)$ to represent any function $f(x)$ such that
$\lim_{x\rightarrow \infty}f(x)/x=0$.

\section{DoF of $2^{L+1}$ X Networks}

\begin{figure}[!h] \vspace{-0.15in}\centering
\includegraphics[width=3.2in]{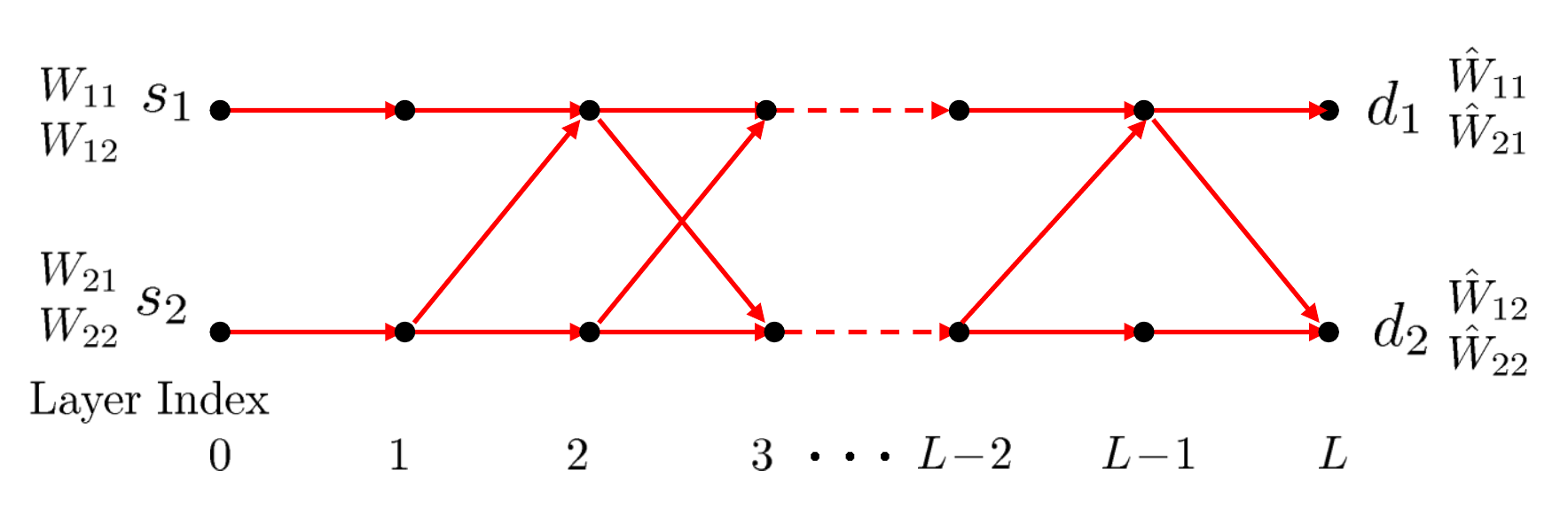}\vspace{-0.15in}
\caption{The $2^{L+1}$ Wireless $X$ network}\vspace{-0.1in}
\label{fig:2_n_network}
\end{figure}

In this section we consider a special class of layered $X$ networks
-- the $2^{L+1}$ $X$ network. By $2^{L+1}$ $X$ network, we mean a
layered multihop $X$ network with $L+1$ layers ($L$ hops), and with
only two nodes at each layer. Such an example is shown in Fig.
\ref{fig:2_n_network}.

\begin{figure}[!h] \vspace{-0.1in}\centering
\includegraphics[width=3in]{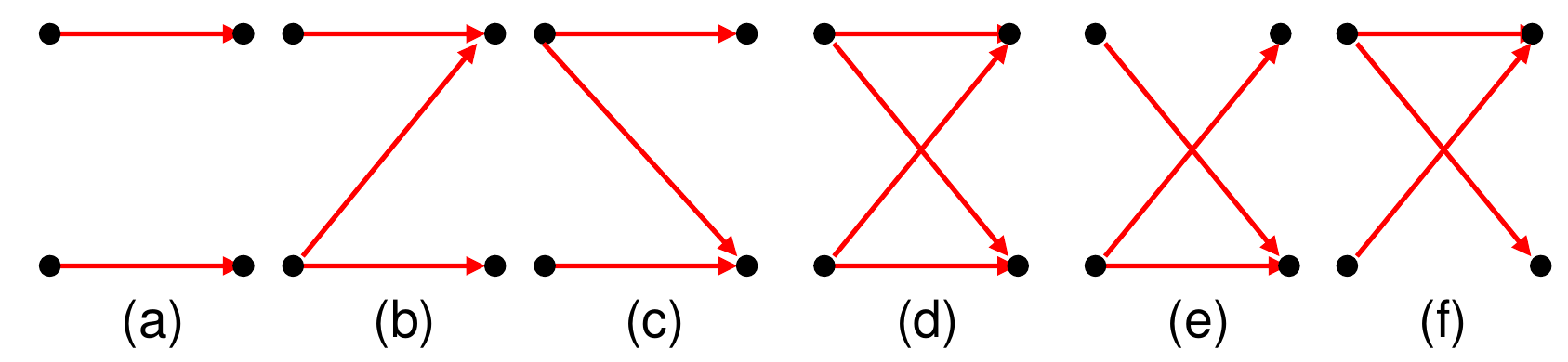}\vspace{-0.15in}
\caption{Components Enumeration of the Wireless $X$ Networks}
\vspace{-0.1in}\label{fig:2_n_component}
\end{figure}

Since the DoF min-cut 1 case is
trivial, let us consider DoF min-cut 2 networks. Between two adjacent layers, we enumerate all topologies of a
one-hop component in Fig. \ref{fig:2_n_component}.
There are six cases (a) to (f) depending on the connectivity. Case
(a) is trivial because two layers can be collapsed into one. Since
the sum capacity of $X$ networks is not affected by switching node
labels within a layer, it is clear that  subnetworks (e) and (b) are
equivalent, and similarly subnetwork (f) is equivalent to (c).
Therefore, in the following we only consider the permutations of
three components (b), (c), (d). For brevity, we call the three
components (b), (c), (d) the ``Z", ``S" and ``X" components,
respectively.

We have the following DoF result.

{\bf Theorem 1}: {\em For  $2^{L+1}$ $X$ networks defined above, the DoF  are given by:
\begin{itemize}
\item[(A)] DoF $=1$, if $L=1$, and the network is a ``Z" or ``S" network.
\item[(B)] DoF $=4/3$, if $L=1$, and the network is an ``X" network.
\item[(C)] DoF $=3/2$, if $L\geq 2$, and the network is one of the eight networks: XZ$^{L-1}$, XS$^{L-1}$, Z$^{L-1}$X, S$^{L-1}$X, ZS$^{L-1}$, SZ$^{L-1}$, S$^{L-1}$Z and Z$^{L-1}$S.
\item[(D)] DoF $=2$, otherwise.
\end{itemize}}

{\it Proof}: {\noindent{Cases (A) and (B)}} follow from previously
known results \cite{Jafar_Shamai}.

{\noindent{\bf Case (C)}:} Since switching node labels within each
layer  does not affect sum-capacity for the $X$ setting, the eight
connectivity patterns for Case (C) can be  reduced to the four
patterns : XZ$^{L-1}$, Z$^{L-1}$X, ZS$^{L-1}$ and S$^{L-1}$Z.
Further, due to the space limitation, we only sketch the argument
that shows that the DoF of XZ$^{L-1}$ network is $\frac{3}{2}$.
Detailed proofs for all connectivity patterns are presented in the
full paper \cite{Wang_Gou_Jafar_MHX}.

\begin{figure}[!h] \vspace{-0.1in}\centering
\includegraphics[width=3in]{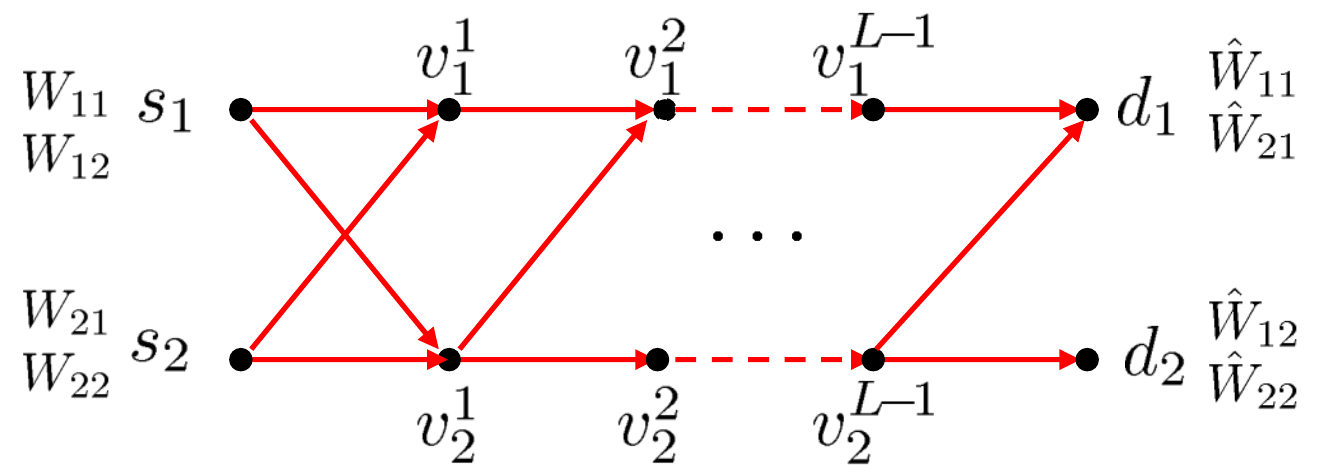}\vspace{-0.15in}
\caption{The XZ$^{L-1}$ Wireless $X$ Network}
\vspace{-0.1in}\label{fig:2_n_xzzz}
\end{figure}

The $L+1$ layered XZ$^{L-1}$ network is shown in
Fig. \ref{fig:2_n_xzzz}. Let us first consider the DoF outer bound.

{\bf DoF Outer Bound:} Because each of the source and destination
nodes has only one antenna, it is trivial to see that the following
four inequalities are satisfied:
\begin{eqnarray}
d_{11}+d_{12}&\!\!\!\!\leq\!\!\!\!& 1,\label{eqn:2messages_bound_1}\\
d_{21}+d_{22}&\!\!\!\!\leq\!\!\!\!& 1,\label{eqn:2messages_bound_2}\\
d_{11}+d_{21}&\!\!\!\!\leq\!\!\!\!& 1,\label{eqn:2messages_bound_3}\\
d_{12}+d_{22}&\!\!\!\!\leq\!\!\!\!& 1.\label{eqn:2messages_bound_4}
\end{eqnarray}
In Fig. \ref{fig:2_n_xzzz}, the destination node $d_2$ can decode
its desired messages $W_{12}$ and $W_{22}$. Since the path from
$v_2^1$ to $d_2$, i.e., $P_{v_2^1,v_2^2,\cdots,v_2^{L\!-\!1},d_2}$,
is free of interference, and the two messages $W_{12},~W_{22}$ must
go through this path, every relay node in this path can decode
$W_{12},~W_{22}$ as well.

Consider the node $v_2^1$. Let us set the message $W_{11}=\phi$ to
bound the rates for the remaining three messages. Since $v_2^1$ is
able to decode the message $W_{12}$, after decoding it $v_2^1$ can
remove the signal carrying $W_{12}$ originating from $s_1$, and then
obtain an AWGN channel directly connected to the source $s_2$.
Therefore, subject to the noise distortion which does not affect the
number of DoF\footnote{We use the phrase "subject to noise
distortion"  to indicate the widely used (see e.g.,
\cite{Jafar_Shamai}) DoF outer bound argument whereby reducing noise
at a node by an amount that is SNR independent (and therefore
inconsequential for DoF) allows it to decode a message.}, $v_2^1$
can also decode the messages $W_{21}$ and $W_{22}$. Since
single-antenna node $v_2^1$ is able to decode all the three messages
$W_{12},~W_{21}$ and $W_{22}$, we have the following DoF outer
bound:
\begin{eqnarray}
d_{12}+d_{21}+d_{22}\leq 1.\label{eqn:3messages_bound_1}
\end{eqnarray}
Similarly we can set $W_{21}=\phi$. Again since $v_2^1$ is able to
decode whatever $d_2$ can decode, $v_2^1$ can decode $W_{22}$ first
and then remove the signal carrying it and thus only sees an AWGN
channel directly connected to $s_1$. Subject to the noise distortion
$v_2^1$ can also decode the messages $W_{11}$ and $W_{12}$, and thus
we have another inequality:
\begin{eqnarray}
d_{11}+d_{12}+d_{22}\leq 1.\label{eqn:3messages_bound_2}
\end{eqnarray}
Adding up all inequalities (\ref{eqn:2messages_bound_3}),
(\ref{eqn:3messages_bound_1}) and (\ref{eqn:3messages_bound_2}), we
have:
\begin{eqnarray}
2(d_{11}+d_{12}+d_{21}+d_{22})\leq 3.
\end{eqnarray}
Therefore, we have the outer bound DoF $\leq \frac{3}{2}$.

\begin{figure}[!h] \vspace{-0.15in}\centering
\includegraphics[width=2.8in]{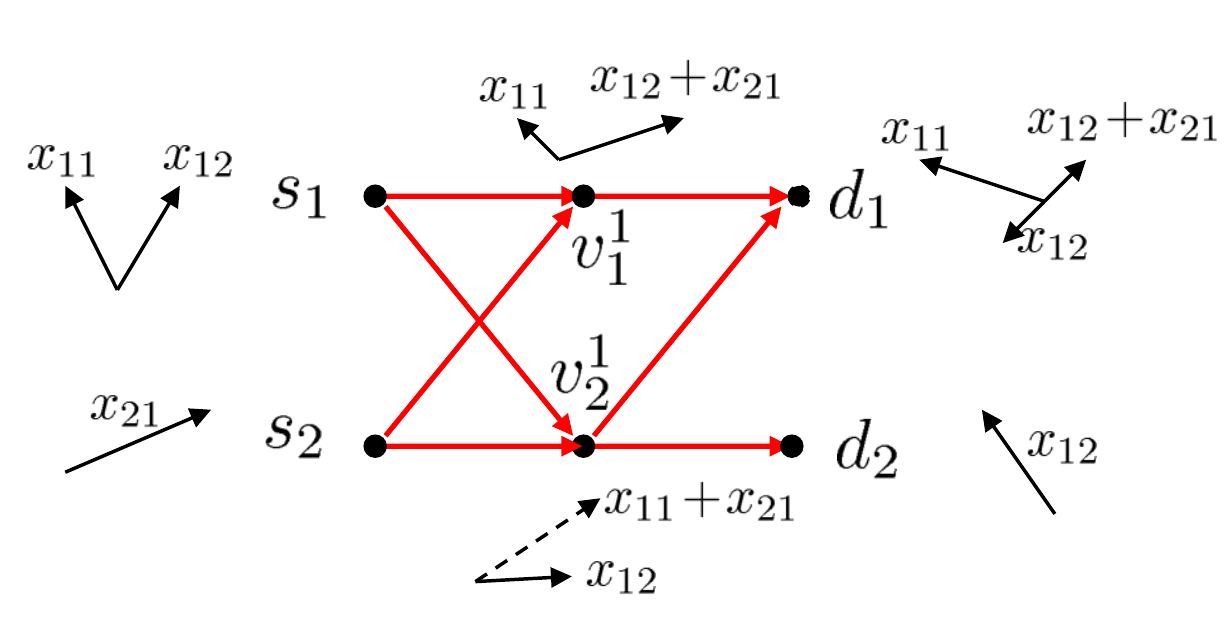}\vspace{-0.15in}
\caption{The ``XZ" Wireless $X$ Network}
\vspace{-0.1in}\label{fig:2_n_xz}
\end{figure}

{\bf Achievability:} We are going to show a simple scheme that can
achieve $3/2$ DoF. We claim that the XZ$^{L-1}$ ($L>2$) network can
also achieve $3/2$ DoF if the two-hop XZ network achieves $3/2$ DoF.
Intuitively, this is because by simply repeating (amplify and forward) whatever the intermediate
relays from the $2^{nd}$ to $(L-1)^{th}$ layers receive, we can
convert ``Z$^{L-1}$" to one ``Z" component. Thus, we only need to
prove $3/2$ DoF is achievable for the two-hop XZ network, as shown
in Fig.\ref{fig:2_n_xz}.

The achievable scheme relies on the idea of aligned interference
neutralization. In addition, since the channel is constant, we will
use signalling in rational dimensions first introduced in
\cite{Etkin_Ordentlich_rational, Motahari_Gharan_Khandani_real}.
Over two rational dimensions\footnote{The concepts of rational
dimensions and rational independence, first proposed for real-valued
numbers, can be applied to complex-valued numbers as well, as
reported in \cite{Maddah-Ali_compoundbc}. Also, if the channel is
time-varying or frequency-selective, we can also create linear space
by using symbol extension, and the rational alignment scheme can be
translated to the linear scheme.}, $s_1$ sends two symbols
$x_{11},~x_{12}$, and $s_2$ sends $x_{21}$ each carrying
$\frac{1}{2}$ DoF and along a ``beamforming'' direction. As shown in
Fig. \ref{fig:2_n_xz} we first randomly pick the direction of
$x_{21}$ at $s_2$. Note that although we use vectors to denote the
``beamforming'' directions in Fig. \ref{fig:2_n_xz} for simplicity,
they should be rationally independent numbers. The direction of
$x_{11}$ at $s_1$ is then fixed by aligning two symbols $x_{11}$ and
$x_{21}$ at $v_2^1$, and the direction of $x_{12}$ at $s_1$ is fixed
by aligning two symbols $x_{12}$ and $x_{21}$ at $v_1^1$. At the
first layer, $v_1^1$ sends two symbols $x_{11}$ and $x_{12}+x_{21}$,
each carried by a randomly picked beamforming direction. The node
$v_2^1$ first demodulates $x_{12}$ and $x_{11}+x_{21}$, and then
only sends $x_{12}$ with a beamforming direction such that $x_{21}$
can be canceled at $d_1$ by the signal coming from $v_1^1$. Since
the directions carrying $x_{11}$ and $x_{21}$ are rationally
independent, the destination $d_1$ is able to decode them, thus
achieving one DoF. Also, because $v_2^1$ only sends $x_{12}$, and
the destination $d_2$ sees a clean channel, $d_2$ can decode its
desired symbol as well to achieve $1/2$ DoF. Therefore, a total of
$3/2$ DoF is achievable.

{\noindent{\bf Case (D)}:} In this case, we claim that except for
the connectivity patterns covered in case (C), all the other
channels where $L\geq 2$ have $2$ DoF. The DoF outer bound for these
networks is trivial and the achievability can be shown based on
eliminating two of the 4 messages to reduce the network to an
interference channel, so that the results of
\cite{Shomorony_Avestimehr} can be applied. The main observation
here, as also in \cite{Shomorony_Avestimehr}, is that in layered
multihop networks if interference arrives through more than one
path, it can be neutralized.

The eight cases in class (C) have two characteristics: (1) there is
only one path from $s_1$ to $d_2$ or from $s_2$ to $d_1$, and (2)
the ``Z" or ``S" components should be consecutive.

If the first condition does not hold, then  it implies that from
$s_1$ to $d_2$ and from $s_2$ to $d_1$, the number of paths is zero
or more than one. In this case, by setting $W_{12}=W_{21}=\phi$,
either there is no interference, or there are more than 1 paths
carrying interference which allows interference neutralization. In
either case, 2 DoF are achieved.

\begin{figure}[!h] \vspace{-0.15in}\centering
\includegraphics[width=2.5in]{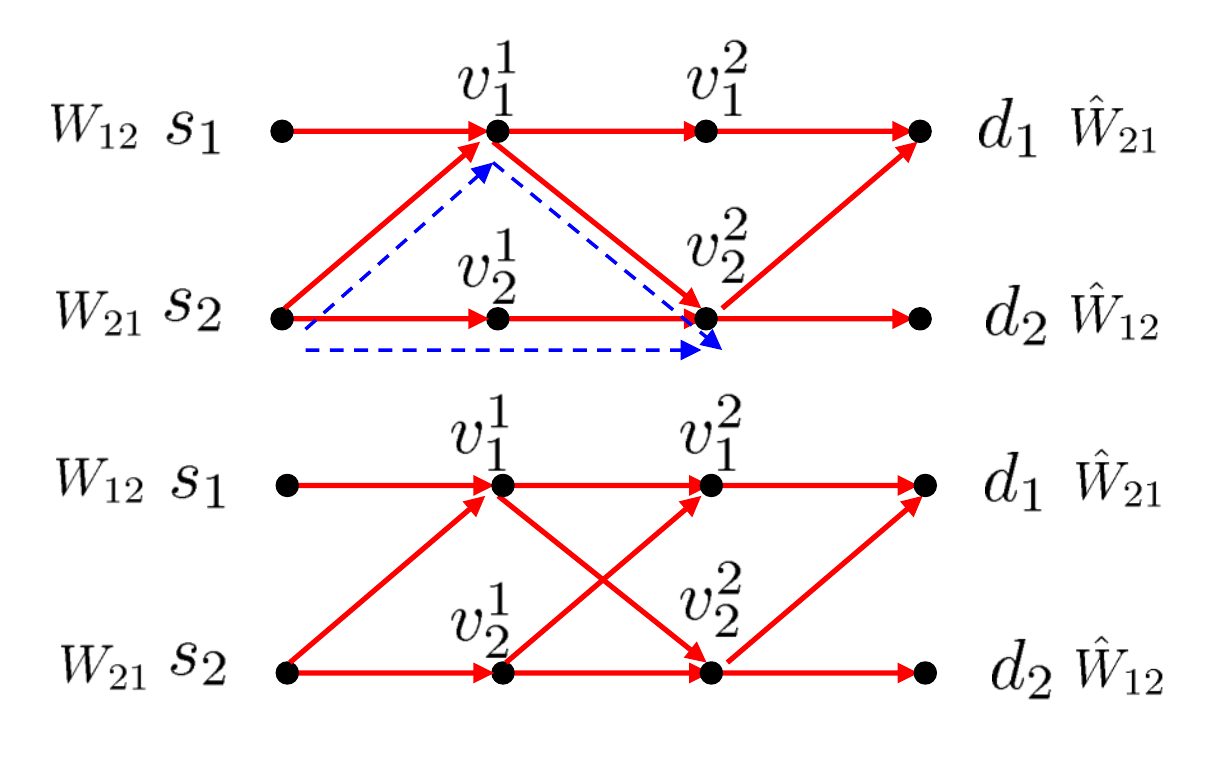}\vspace{-0.15in}
\caption{$X$ Networks with Two DoF} \vspace{-0.1in}
\label{fig:2_n_zsz_zxz}
\end{figure}

Next, consider the class of connectivity patterns where the second
condition is not satisfied but the first condition still holds. In
Fig. \ref{fig:2_n_zsz_zxz} we show two specific networks
characterizing the main properties of this class of networks. Take
the first as an example, if we set $W_{11}=W_{22}=\phi$ then it
forms a two user interference network, in which it is easy to see
that two DoF are achievable even though there is only one path from
$s_1$ to $d_2$. This is because the message $W_{21}$ intended for
the destination $d_1$ can always be nulled at $v_2^2$ after going
through two paths $P_{s_2,v_1^1,v_2^2}$ and $P_{s_2,v_2^1,v_2^2}$
(denoted by the dashed lines) such that $v_2^2$ and thus $d_2$ is
interference-free. Since $W_{12}$ can still arrive at $d_2$ through
the path $P_{s_1,v_1^1,v_2^2,d_2}$, the destination $d_2$ can
achieve one DoF. Similarly, $d_1$ can also ahieve one DoF. Thus, a
total of two DoF is achievable. \hfill\QED



%

\section{DoF of General Multihop Layered $X$ Network}
So far, we studied the DoF of $X$ networks where there are only two
relay nodes at each layer. If the number of relay nodes is not
limited to two, one question is whether the DoF of the network with
arbitrary connectivity still belong to the set
$\{1,\frac{4}{3},\frac{3}{2},2\}$. Interestingly, there is another
class of $X$ networks that have $5/3$ DoF. Due to the space
limitation, we will only show one specific example in such a new
class. The detailed description and analysis for this new class are
reported in \cite{Wang_Gou_Jafar_MHX}.

\begin{figure}[!h] \vspace{-0.15in}\centering
\includegraphics[width=3.5in]{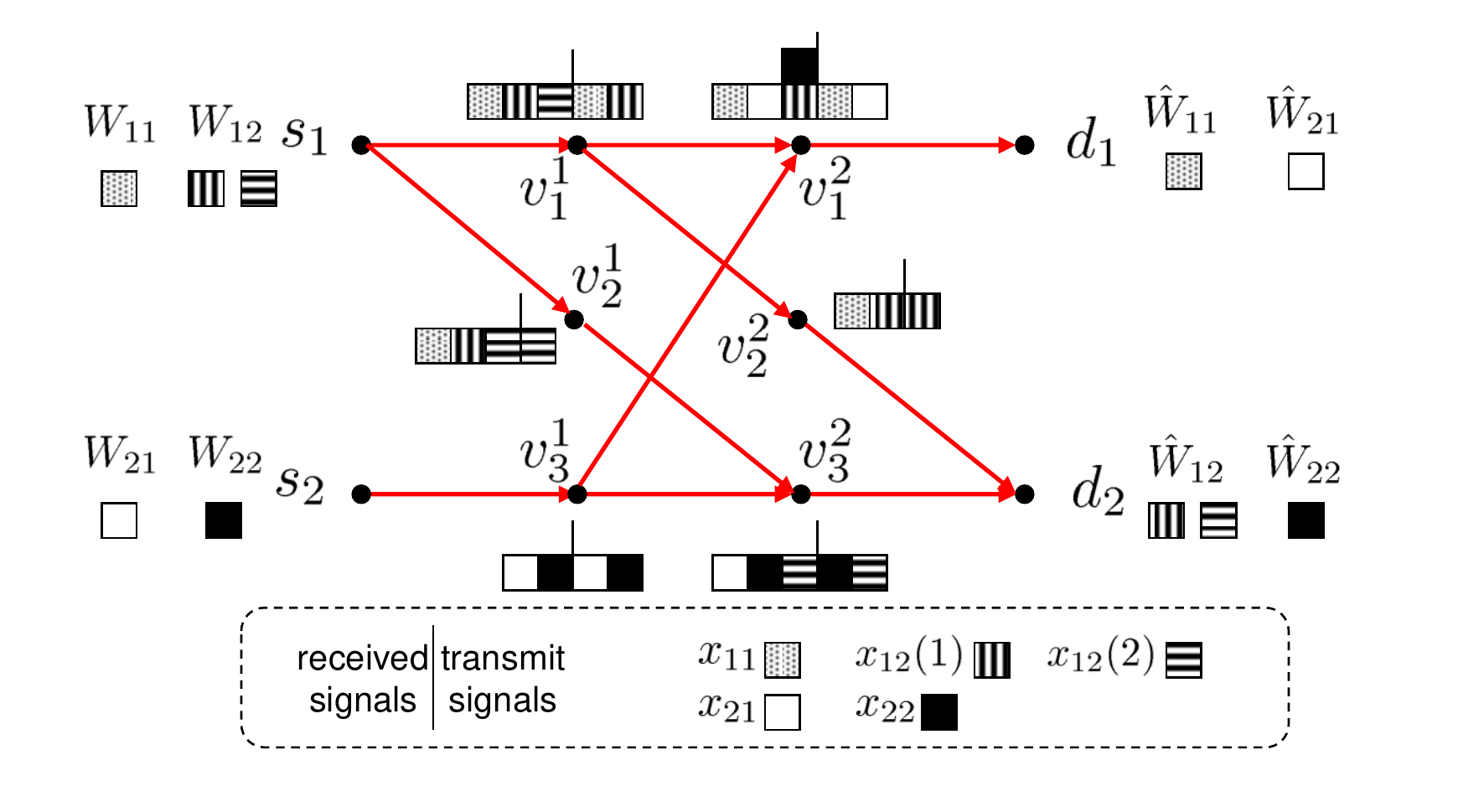}\vspace{-0.2in}
\caption{A Three-hop Layered Network wth $5/3$ DoF}\vspace{-0.1in}
\label{fig:5_over_3_network}
\end{figure}

{\bf Theorem 2}: The network of Fig. \ref{fig:5_over_3_network} has $5/3$ DoF.




\subsection{DoF Outer Bound}

Similar to the analysis in Case (C) of Section III, if we set
$W_{12}=\phi$, then we have the following inequality:
\begin{eqnarray}
d_{11}+d_{21}+d_{22}\leq 1.\label{eqn:out_5over3_2}
\end{eqnarray}
Notice that $Y^n_{d_1}$ only depends on $Y^n_{v_1^2}$. Because $d_1$
is able to decode $W_{11}$, $v_1^2$ can decode it as well. After
decoding $W_{11}$, $v_1^2$ can subtract the signal $X^n_{v_1^1}$
carrying $W_{11}$, thus obtaining $X_{v_3^1}^n$ subject to the noise
distortion which depends on the channel coefficients but is
independent of SNR, and thus has only an $o(\log(\rm SNR))$ impact
on the rate. Since $v_3^1$ can decode $W_{21}$ and $W_{22}$, $v_1^2$
can decode $W_{21},W_{22}$ as well. Thus, the single-antenna node
$v_1^2$ can decode all these three messages, which implies
(\ref{eqn:out_5over3_2}).

Next we derive a new information-theoretical DoF outer
bound. Consider the sum rate of two messages desired at the destination
$d_1$. A genie provides $W_{22}$ to the node $v_1^2$.
\allowdisplaybreaks
\begin{eqnarray}
&\!\!\!\!\!\!\!\!&n(R_{W_{21}}+R_{W_{11}})\notag\\
&\!\!\!\!\leq\!\!\!\!& I(W_{21},W_{11};Y_{d_1}^n)+o(n)\label{eqn:out_1_fano}\\
&\!\!\!\!\leq\!\!\!\!& I(W_{21},W_{11};Y_{v_1^2}^n)+o(n)\label{eqn:out_1_dpt}\\
&\!\!\!\!\leq\!\!\!\!& I(W_{21},W_{11};Y_{v_1^2}^n,W_{22})+o(n)\label{eqn:out_1_genie}\\
&\!\!\!\!\leq\!\!\!\!& I(W_{21},W_{11};W_{22})+I(W_{21},W_{11};Y_{v_1^2}^n|W_{22})+o(n)\label{eqn:out_1_chain}\\
&\!\!\!\!=\!\!\!\!& I(W_{21},W_{11};Y_{v_1^2}^n|W_{22})+o(n)\label{eqn:out_1_indep}\\
&\!\!\!\!=\!\!\!\!&h(Y_{v_1^2}^n|W_{22})-h(Y_{v_1^2}^n|W_{21},W_{11},W_{22})+o(n)\\
&\!\!\!\!\leq\!\!\!\!& n(\log\rho)-h(Y_{v_1^2}^n|W_{21},W_{11},W_{22})+o(n)\\
&\!\!\!\!=\!\!\!\!&n(\log\rho)-h(Y_{v_2^2}^n|W_{21},W_{11},W_{22})\!+\!n~o(\log\rho)\!+\!o(n\!).\label{eqn:out_1}\
\ \ \ \ \
\end{eqnarray}

Here, (\ref{eqn:out_1_fano}) follows from Fano's inequality.
(\ref{eqn:out_1_dpt}) follows from the data processing inequality,
because $(W_{11},W_{21})-Y_{v_1^2}^n-Y_{d_1}^n$ forms a Markov
chain. (\ref{eqn:out_1_genie}) is obtained because providing genie
does not decrease the capacity region of the network.
(\ref{eqn:out_1_chain}) follows from the chain rule.
(\ref{eqn:out_1_indep}) is obtained since $W_{22}$ is independent of
$(W_{21},W_{11})$. (\ref{eqn:out_1}) follows from the invertibility
of the channels (regardless of the values of the channel
coefficients as long as they are all non-zero), which implies that
given $(W_{21},W_{11},W_{22})$ the entropy of $Y_{v_2^2}^n$ is equal
to that of $Y_{v_1^2}^n$ subject to the noise distortion.
Specifically, knowing $(W_{21},W_{22})$ we can reconstruct the
signal $X_{v_3^1}^n$, and by subtracting it from $Y_{v_1^2}^n$ we
obtain the signal $X^n_{v_1^1}$ subject to the noise distortion
which will depends on the channel coefficients but is independent of
SNR. The entropy of $Y_{v_2^2}^n$ is equal to that of $X_{v_1^1}^n$
subject to the noise distortion which again depends only on the
channel coefficients but is independent of SNR. All these operations
only have an $o(\log({\rm SNR}))$ impact on rate, and so we obtain
$h(Y_{v_1^2}^n|W_{21},W_{11},W_{22})=h(Y_{v_2^2}^n|W_{21},W_{11},W_{22})+n~o(\log\rho)$
as shown in (\ref{eqn:out_1}). By rearranging terms of
(\ref{eqn:out_1}) we obtain the first outer bound:
\begin{small}
\begin{eqnarray}
\vspace{-0.125in}\mbox{$n(R_{W_{21}}\!+\!R_{W_{11}})\!+\!h(Y_{v_2^2}^n|W_{21},W_{11},W_{22})\leq$}\ \ \ \ \ \ \ \ \ \ \ \ \ \ \ \notag\\
\hspace{-0.125in} \ \ \ \ \ \ \ \ \ \ \ \ \ \ \ \ \ \ \ \
\mbox{$n(\log \rho)+n~o(\log\rho)+o(n)$}.\label{eqn:out_1_rearrange}
\end{eqnarray}
\end{small}\vspace{-0.2in}

Next, let us consider the sum rate of two messages originating from
the source $s_2$. A genie provides $W_{11}$ to the nodes $v_2^2$ and
$v_3^2$. \vspace{-0.05in}
\begin{eqnarray}
\vspace{-0.125in}&\!\!\!\!\!\!\!\!&\mbox{$n(R_{W_{21}}+R_{W_{22}})$}\notag\\
&\!\!\!\!\leq\!\!\!\!& I(W_{21},W_{22};Y^n_{d_1},Y^n_{d_2})+o(n)\label{eqn:out_2_fano}\\
&\!\!\!\!\leq\!\!\!\!& I(W_{21},W_{22};Y^n_{v_1^2},Y^n_{v_2^2},Y^n_{v_3^2})+o(n)\label{eqn:out_2_dpt}\\
&\!\!\!\!\leq\!\!\!\!& I(W_{21},W_{22};Y^n_{v_1^2},Y^n_{v_2^2},Y^n_{v_3^2},{W_{11}})+o(n)\label{eqn:out_2_genie}\\
&\!\!\!\!=\!\!\!\!& I(W_{21},W_{22};Y^n_{v_2^2},Y^n_{v_3^2},{W_{11}})\notag\\
&\!\!\!\!\!\!\!\!& +I(W_{21},W_{22};Y^n_{v_1^2}|Y^n_{v_2^2},Y^n_{v_3^2},{W_{11}})+o(n)\label{eqn:out_2_chain}\ \ \ \\
&\!\!\!\!\leq\!\!\!\!& I(W_{21},W_{22};Y^n_{v_2^2},Y^n_{v_3^2},{W_{11}})+n~o(\log\rho)+o(n)\ \ \ \ \label{eqn:out_2_recover}\\
&\!\!\!\!=\!\!\!\!& I(W_{21},W_{22};Y^n_{v_3^2}|Y^n_{v_2^2},{W_{11}})\notag\\
&\!\!\!\!\!\!\!\!& +I(W_{21},W_{22};Y^n_{v_2^2},{W_{11}})+n~o(\log\rho)+o(n)\\
&\!\!\!\!=\!\!\!\!& I(W_{21},W_{22};Y^n_{v_3^2}|Y^n_{v_2^2},{W_{11}})+n~o(\log\rho)+o(n)\label{eqn:out_2_indpt}\\
&\!\!\!\!\leq\!\!\!\!& h(Y^n_{v_3^2}|Y^n_{v_2^2},{W_{11}})+n~o(\log\rho)+o(n)\notag\\
&\!\!\!\!\!\!\!\!& -h(Y^n_{v_3^2}|Y^n_{v_2^2},{W_{11}},{W_{21}},{W_{22}})\label{eqn:out_2_h_chain}\\
&\!\!\!\!\leq\!\!\!\!&
n(\log\rho)-h(Y^n_{v_3^2}|Y^n_{v_2^2},W_{11},W_{21},W_{22})\notag\\
&\!\!\!\!\!\!\!\!&+n~o(\log\rho)+o(n)\label{eqn:out_2}
\end{eqnarray}
where (\ref{eqn:out_2_fano}) follows from Fano's inequality.
(\ref{eqn:out_2_dpt}) follows from the data processing inequality
because $(W_{21},W_{22})-Y_{v_1^2}^n-Y_{d_1}^n$, and
$(W_{21},W_{22})-(Y^n_{v_2^2},Y^n_{v_3^2})-Y_{d_2}^n$ form two
Markov chains. (\ref{eqn:out_2_recover}) follows from the
invertibility of channels which implies that using
$(Y^n_{v_2^2},Y^n_{v_3^2},{W_{11}})$ we can recover the signal
$Y^n_{v_1^2}$ subject to the noise distortion. Specifically, we can
use $(Y^n_{v_2^2},Y^n_{v_3^2})$ to decode $W_{12}$ because
$Y^n_{d_2}$ only depends on $(Y^n_{v_2^2},Y^n_{v_3^2})$. Thus, using
${W_{11}}$ and ${W_{12}}$ we can recover the signal $X^n_{s_1}$
subject to the noise distortion. By knowing $X^n_{s_1}$ we also know
$(X^n_{v_1^1},X^n_{v_2^1})$. Thus, given
$(X^n_{v_2^1},Y^n_{v_3^2})$, we can reconstruct the signal
$X^n_{v_3^1}$ subject to the noise distortion. Finally, given
$(X^n_{v_3^1},X^n_{v_1^1})$ we can reconstruct the signal
$Y^n_{v_1^2}$. All these operations only have an $o(\log({\rm
SNR}))$ impact on rate, so we obtain
$I(W_{21},W_{22};Y^n_{v_1^2}|Y^n_{v_2^2},Y^n_{v_3^2},{W_{11}})\leq
n~o(\log\rho)+o(n)$ as in (\ref{eqn:out_2_recover}).
(\ref{eqn:out_2_indpt}) is obtained since $(W_{21},W_{22})$ are
independent of $(Y^n_{v_2^2},{W_{11}})$. (\ref{eqn:out_2_h_chain})
follows from the chain rule. By rearranging terms of
(\ref{eqn:out_2}) we obtain the second outer bound:
\begin{small}\vspace{-0.05in}
\begin{eqnarray}
\hspace{-0.125in} \mbox{$n(R_{W_{21}}\!+\!R_{W_{22}}\!)+h(Y^n_{v_3^2}|Y^n_{v_2^2},{W_{11}},{W_{21}},{W_{22}})\leq$}\ \ \ \ \ \ \ \ \notag\\
\hspace{-0.125in} \ \ \ \ \ \ \ \ \ \ \ \ \ \ \ \
\mbox{$n(\log\rho)+n~o(\log\rho)+o(n)$}.\label{eqn:out_2_rearrange}
\end{eqnarray}
\end{small}

In the following, we consider the rate of message $W_{12}$. We
provide genie $(W_{11},W_{21},W_{22})$ to the nodes $(v_2^2,v_3^2)$.
\begin{eqnarray}
nR_{W_{12}}\!\!&\!\!\!\!\leq\!\!\!\!& I(W_{12};Y^n_{d_2})+o(n)\label{eqn:out_3_fano}\\
&\!\!\!\!\leq\!\!\!\!& I(W_{12};Y^n_{v^2_2},Y^n_{v^2_3})+o(n)\label{eqn:out_3_dpt}\\
&\!\!\!\!\leq\!\!\!\!& I(W_{12};Y^n_{v^2_2},Y^n_{v^2_3},W_{11},W_{21},W_{22})+o(n)\label{eqn:out_3_genie}\\
&\!\!\!\!\leq\!\!\!\!& I(W_{12};W_{11},W_{21},W_{22})+o(n)\notag\\
&\!\!\!\!\!\!\!\!&+I(W_{12};Y^n_{v^2_2},Y^n_{v^2_3}|W_{11},W_{21},W_{22})\label{eqn:out_3_chain}\\
&\!\!\!\!=\!\!\!\!& h(Y^n_{v^2_2},Y^n_{v^2_3}|W_{11},W_{21},W_{22})+o(n)\notag\\
&\!\!\!\!\!\!\!\!&-h(Y^n_{v^2_2},Y^n_{v^2_3}|W_{11},W_{12},W_{21},W_{22})\\
&\!\!\!\!\leq\!\!\!\!&
h(Y^n_{v^2_2},Y^n_{v^2_3}|W_{11},W_{21},W_{22})\!+n~o(\log\rho)\!+\!o(n\!)\label{eqn:out_3_indpt}\ \ \ \ \ \\
&\!\!\!\!\leq\!\!\!\!&
h(Y^n_{v^2_2}|W_{11},W_{21},W_{22})+n~o(\log\rho)+o(n)\notag\\
&\!\!\!\!\!\!\!\!&+h(Y^n_{v^2_3}|Y^n_{v^2_2},W_{11},W_{21},W_{22})\label{eqn:out_3}
\end{eqnarray}
where (\ref{eqn:out_3_fano}) follows from Fano's inequality.
(\ref{eqn:out_3_indpt}) is obtained because knowing the four
messages $(W_{11},W_{12},W_{21},W_{22})$ we can reconstruct the
signals $(Y^n_{v^2_2},Y^n_{v^3_2})$ subject to the noise distortion.

Adding up inequalities (\ref{eqn:out_1_rearrange}),
(\ref{eqn:out_2_rearrange}) and (\ref{eqn:out_3}), we have:
\begin{small}
\begin{eqnarray}
n(2R_{W_{21}}\!\!+\!\!R_{W_{11}}\!\!+\!\!R_{W_{22}}\!\!+\!\!R_{W_{12}}\!)\!\leq\!
2n(\log\!\rho)\!+\!n~o(\log\!\rho)\!+\!o(n).\!\!\!\label{eqn:out_5over3_0}
\end{eqnarray}
\end{small}
Dividing $n(\log\rho)$ on both sides of (\ref{eqn:out_5over3_0}),
and taking $n\rightarrow \infty$, $\rho\rightarrow \infty$, we
obtain the following inequality:
\begin{eqnarray}
2d_{21}+d_{11}+d_{22}+d_{12}\leq 2.\label{eqn:out_5over3_1}
\end{eqnarray}
Now adding up inequalities (\ref{eqn:2messages_bound_1}),
(\ref{eqn:2messages_bound_4}), (\ref{eqn:out_5over3_2}) and
(\ref{eqn:out_5over3_1}), we obtain:
\begin{eqnarray}
3(d_{11}+d_{12}+d_{21}+d_{22})\leq 5.
\end{eqnarray}
Thus, the total DoF of this network is bounded above by $5/3$.

\subsection{Achievability of $5/3$ DoF}
We provide an interference alignment scheme that can achieve $5/3$
DoF in the network in Fig.\ref{fig:5_over_3_network}. Over three
rational dimensions, source $s_1$ sends one symbol $x_{11}$ to
$d_1$, two symbols $x_{12}(1),~x_{12}(2)$ to $d_2$, and $s_2$ sends
one symbol $x_{21}$ to $d_1$, and one symbol $x_{22}$ to $d_2$, each
carrying $\frac{1}{3}$ DoF along a rationally independent
``beamforming'' direction. For brevity, in Fig.
\ref{fig:5_over_3_network} we also use  boxes (each box denotes one
symbol, carrying $1/3$ DoF) with different patterns, to show how our
scheme works. We consider the transmission schemes from each layer
to the next in what follows.


{\bf From layer 0 to layer 1:} The source $s_1$ randomly picks three
rationally independent beamforming directions to transmit
$x_{11},~x_{12}(1),~x_{12}(2)$. Because the edges $P_{s_1,v_1^1}$
and $P_{s_1,v_2^1}$ are AWGN channels, $v_1^1$ and $v_2^1$ both can
decode these three symbols. Similarly, $v_3^1$ can decode
$x_{21},~x_{22}$.

{\bf From layer 1 to layer 2:} After decoding $x_{11}$, $x_{12}(1)$
and $x_{12}(2)$, $v_1^1$ sends $x_{11}$ and $x_{12}(1)$ to $v_1^2$
and $v_2^2$ using any two rationally independent beamforming
directions $U_{v_1^1}(x_{11})$ and $U_{v_1^1}(x_{12}(1))$,
respectively. $v_2^1$ only sends $x_{12}(2)$ to $v_3^2$ with a
randomly picked beamforming direction. $v_3^1$ sends $x_{21}$ with
another randomly picked beamforming direction, but $x_{22}$ in the
direction $U_{v_3^1}(x_{22})$ such that $x_{22}$ aligns with
$x_{12}(1)$ at $v_1^2$ in the same dimension, i.e., $H_{v_1^2
v_3^1}U_{v_3^1}(x_{22})=H_{v_1^2 v_1^1}U_{v_1^1}(x_{12}(1))$.

{\bf From layer 2 to layer 3:} The node $v_1^2$ can see a
three-rational dimensional space, each dimension carrying symbols
$x_{11}$, $x_{21}$ and $x_{12}(1)+x_{22}$, respectively. Thus, it
can demodulate these symbols and only transmits $x_{11}$, $x_{21}$
to the destination $d_1$ such that $d_1$ achieves $2/3$ DoF.
$v_2^2$ receives two symbols $x_{11}$ and $x_{12}(1)$ in two
dimensions, and thus it can demodulate them and only sends
$x_{12}(1)$ to the destination $d_2$ with an randomly picked
beamforming direction. Similarly, $v_3^2$ receives three symbols
$x_{21},~x_{22},~x_{12}(2)$ in three rationally independent
dimensions. Thus, it can demodulate them and only transmits
$x_{22},~x_{12}(2)$ to $d_2$ with two randomly picked rationally
independent beamforming directions. At the destination $d_2$,
because it receives three desired symbols in three rationally
independent dimensions, it can achieve 1 DoF.

Therefore, a total of $5/3$ DoF is achievable almost surely.

Since both outer and inner bounds are $\frac{5}{3}$ DoF, we
establish that the network has a total of  $5/3$ DoF. \hfill\QED

\section{Conclusion}

Total degrees of freedom (DoF) for multiple unicasts over 2 source 2
sink layered multihop wireless networks are shown to take values $1,
4/3, 3/2, 2$, depending on the connectivity within each hop, for
almost all values of channel coefficients, when the number of
relayed in each layer is no more than 2. If the number of relays at
each layer is not restricted to 2, it is shown through an example,
that the network can also have DoF value $5/3$. Finally, we are able
to show in \cite{Wang_Gou_Jafar_MHX} that 1, 4/3, 3/2, 5/3 and 2 (in
the almost surely sense) are the only possible DoF values  for {\em
all} connectivity patterns.


\section*{Acknowledgement}
The authors would like to acknowledge prior
collaboration with Hamed Maleki at UC Irvine and Prof. Sriram
Vishwanath at UT Austin for the wired $X$ network capacity result.

\newpage


\begin{thebibliography}{1}

\bibitem{Gou_Jafar_222}
T.~Gou, S.~Jafar, S.~Jeon, S.~Chung, ``Aligned Interference
Neutralization and the Degrees of Freedom of the $2\times 2\times 2$
Interference Channel", {\em e-print arXiv:1012.2350}, Dec. 2010.

\bibitem{Shomorony_Avestimehr}
I.~Shomorony and S.~Avestimehr, ``Two unicast wireless networks:
characterizing the degrees-of-freedom'', \emph{arXiv:1102.2498},
Mar. 2011.


\bibitem{Cai_Letaief_Fan_Feng}
K. Cai, K. B. Letaief, P. Fan, R. Feng, ``On the Solvability of
2-pair Unicast Networks --- A Cut-based Characterization", {\em
arXiv:1007.0465v1}, July 2010.


\bibitem{Jafar_Shamai}
S.~Jafar, S.~Shamai, ``Degrees of Freedom Region for the MIMO $X$
Channel'', {\em IEEE Transactions on Information Theory}, vol.~54,
No.~1, pp.~151-170, Jan. 2008.

\bibitem{Cadambe_Jafar_X}
V.~Cadambe and S.~Jafar, ``Interference alignment and the degrees of
freedom of wireless $X$ networks'', {\em IEEE Trans. on Information
Theory}, vol.~55, no. 9,  pp.~3893--3908, Sep. 2009.

\bibitem{MMK}
M.A. Maddah-Ali, A.S. Motahari , and A.K. Khandani, "Communication
Over MIMO $X$ Channels: Interference Alignment, Decomposition, and
Performance Analysis," IEEE Transaction on Information Theory, Vol.
54, No. 8, pp. 3457-3470, Aug. 2008.


\bibitem{Wang_Gou_Jafar_MHX}
C.~Wang, T.~Gou and S.~Jafar, ``Degrees of Freedom of Layered
Multihop $X$  Networks'', {\em Full paper in preparation.}

\bibitem{Etkin_Ordentlich_rational}
R. Etkin and E. Ordentlich, ``The Degrees of Freedom of the K User
Gaussian Interference Channel Is Discontinuous at Rational Channel
Coefficients", {\em IEEE Trans. on Infor. Theory}, vol. 55, no. 11,
Nov. 2009.

\bibitem{Motahari_Gharan_Khandani_real}
A.S. Motahari, S. O. Gharan and A. K. Khandani, ``Real Interference
Alignment with Real Numbers," {\em arXiv:0908.1208}, Aug. 2009.

\bibitem{Maddah-Ali_compoundbc}
M.A. Maddah-Ali, ``On the Degrees of Freedom of the Compound MIMO
Broadcast Channels with Finite States", {\em arXiv:0909.5006v3},
Oct. 2009.


%







\end{thebibliography}
\end{document}